\documentclass[onecolumn,showpacs,floatfix,11pt,nofootinbib]{revtex4}
\usepackage[dvips]{graphics,color}
\usepackage{epsfig}\usepackage{float}
\RequirePackage{amssymb}
\usepackage{makeidx}
\usepackage{graphicx}
\usepackage{indentfirst}
\usepackage{color}
\newcommand{\be}{\begin{equation}}
\newcommand{\ee}{\end{equation}}

\newcommand{\ba}{\begin{eqnarray}}
\newcommand{\ea}{\end{eqnarray}}

\newcommand{\el}{^}
\begin{document}

\title{Effective gravitational equations for $f(R)$ braneworld models}
\author{T. R. P. Caram\^es}
\email{carames@if.uff.br}
\author{M. E. X. Guimar\~aes}
\email{emilia@if.uff.br}
\affiliation{Instituto de F\'isica,
Universidade Federal Fluminense\\
Av. Gal. Milton Tavares de Souza, s/n \\
CEP 24210-346, Niter\'oi-RJ, Brazil.}
\author{J. M. Hoff da Silva}
\email{hoff@feg.unesp.br;hoff@ift.unesp.br}
\affiliation{UNESP - Campus de Guaratinguet\'a - DFQ\\
Av. Dr. Ariberto Pereira da Cunha, 333 \\
CEP 12516-410, Guaratinguet\'a-SP,
Brazil.}


\begin{abstract}
The viability of achieving gravitational consistent braneworld models in the framework of a $f(R)$ theory of gravity is  investigated. After a careful generalization of the usual junction conditions encompassing the embedding of the 3-brane into a $f(R)$ bulk, we provide a prescription giving the necessary constraints in order to implement the projected second order effective field equations on the brane.
\end{abstract}

\maketitle

\section{Introduction}

In the last decade braneworld models performed a genuine branch of research in high energy physics. In particular, the two papers of Randall and Sundrum \cite{RSI,RSII} envisage our universe as a warped brane embedded into a five dimensional spacetime, the bulk. Several extensions and/or modifications of this original set up has been proposed in the literature. These extensions include, for instance, other dimensions on the brane and in the bulk, the possibility of additional fields in the bulk, and the consideration of modified bulk gravity \cite{greg, AGH1, AGH2}. Among the possible modifications in the bulk gravity, the consideration of higher derivative models was extensively analyzed (for comprehensive reviews in $f(R)$ gravity see \cite{REV}). In Ref. \cite{EV} it is shown that $f(R)$ theories may help in avoiding the fine-tuning problem of the Randall-Sundrum \cite{RSI} model. Besides, several interesting cosmological aspects of $f(R)$ braneworld models were investigated \cite{VARI,VARII,VARIII}. It was also demonstrated recently that in the five-dimensional $f(R)$ braneworld models scope it is possible to obtain a model which solves the hierarchy problem without the necessity of a negative tension brane in the compactification scheme \cite{EN, AGH3}.

When dealing with gravitational aspects of braneworld models, one naturally faces the possibility of generalizing the usual field equations on the brane. In fact, even for a bulk respecting solely the General Relativity theory, the projected equations on the brane are modified leading to subtle but important departures from standard scenarios.

Our aim in this work is to obtain the appropriate junction conditions and the effective second field equations on a 3-brane embedded in a five-dimensional $f(R)$ bulk. At this point, some remarks on the previous literature are in the order. The junction conditions for higher-order braneworld models with a Gibbons-Hawking term were founded in the Ref. \cite{AM} and the accordingly cosmological equations for the fourth-order $f(R)$ braneworld were obtained in \cite{AM2}. In these precise and encompassing papers, as well as in \cite{EV}, the junction conditions are obtained by means of the conformal transformation relating $f(R)$ and scalar-tensor theories \cite{ANTIG}. In fact, there are many advantages in the conformal theory since the resulting field equations are again of second order (with matter). As it is well known, the conformal transformation performs an isomorphism between the space of solutions of the two conformally related theories \cite{BAO}, providing a mathematical equivalence between them. Besides, the physical content of the theory must not depend on the conformal frame in which the theory is formulated or presented \cite{REFNOVO}. Therefore, as we shall obtain the effective projected equations for the most general case (with matter on the brane), we prefer to reobtain the junction conditions via the usual brute force method. Only in Sec. III we particularize, for simplicity, our equations to the brane vacuum case and therefore we shall approach the effective field equations in a conservative and general way. Finally, we point out that this program was carried out in Ref. \cite{IR}. Nevertheless in \cite{IR}, and subsequently in \cite{CONT}, the same junction conditions as in the General Relativity case were used, which is, in fact, only a (non rigorous) over simplification, since the junction conditions depends on the embedding gravity space. As we are interested in the effective second order (Einstein like) brane gravitational equations our procedure towards the suitable junction conditions is well defined. We remark, however, that for higher order equations on the brane it is necessary to implement a more powerful method to improve the junction conditions \cite{NR}. Recently, the necessity of the new junction conditions was circumvented in some cosmological applications to the braneworld case \cite{DENO}.

An important remark here is needed to be clarified in order to avoid any misunderstandings: in the ref. \cite{deruelle}, the authors also use the Israel-Darmois junction conditions to study the same problem as here. However, as the authors emphasize in their paper ``when $f(R) = R$ the junction conditions (3.10) do not reduce to the familiar Israel conditions 34) as they have to be supplemented by the condition of continuity of $R$. What happens when $f(R) = -2\Lambda + R+ l^2R^2+...$ when $l \rightarrow 0$  is that the bulk geometry may approach a solution of the Einstein bulk equations (e.g., AdS) everywhere, to the exception of a region of size $ l$  in the vicinity of the brane, so that when $ l$ becomes very small the thin shell limit is no longer valid and the thickness of the brane must be taken into account." Here, we avoid a conformal transformation as in ref. \cite{deruelle} exactly because we do not touch the problem of the brane having a thickness or not. If the brane has a thickness, indeed there would appear a size scale and a conformal transformation would be very difficult to be carried on. The fact that we both do not deal with a brane thickness and do not make any conformal transformation leads us to the constraint equations (32-33) which reduce to the usual Israel-Darmois condition when $ f(R) =R$ (e.g., the GR case) and there is no need to study the continuity of $R$ through the brane. Therefore, though both papers are dealing with the same subject, the approaches are very different.

This paper is structured as follows: in the next Section we briefly review the scalaron in the bulk. In the section III we obtain the junction conditions for the $f(R)$ bulk gravity without using the conformal transformation method. It is shown that in the appropriate limit we recover the usual Israel-Darmois equations. Moving forward we obtain the effective second order projected field equations on the brane. In Sec. IV we implement several necessary constraints in the bulk $f(R)$ gravity and study the correspondent implications on the brane equations in order to set a physically acceptable and viable model. Finally, in the conclusion we summarize our main results and give a simple example of how to apply the general projection to a specific (toy-)model.

\section{Field Equations and the Extra Scalar Degree of Freedom}

An interesting aspect observed in a $f(R)$ theory of gravity is an emerging scalar degree of freedom. This property is resulting straightforwardly from the modification of gravity and allows us to set a dynamical equivalence of these theories with scalar-tensor ones, in particular Brans-Dicke gravity. Such equivalence is verified in both variational formalisms metric and Palatini. Within the metric approach this correspondence has already been widely discussed in the literature \cite{METRIC}. From the viewpoint of the Palatini formalism this feature has been analized in the references \cite{PALAT}.

The gravitational field equations in the metric formalism for a five-dimensional bulk in a $f(R)$ gravity is given by
\be f'(R)R_{AB}-\frac{1}{2}g_{AB}f(R)+g_{AB}\Box f'(R)-\nabla_{A}\nabla_{B}f'(R)=k\el{2}_{5}T_{AB},\label{1}\ee where $A=0,...,4$, $f'(R)=df(R)/dR$ and $k_{5}\el{2}$ is the five-dimensional gravitational coupling constant. The extra scalar degree of freedom mentioned above which arises in $f(R)$ gravity can be observed when we take the trace of the above equation, which reads
\begin{equation}
\label{trace}
f'(R)R-\frac{5}{2}f(R)+4\Box f'(R)=k_{5}^{2}T\ .
\end{equation}
We verify that the trace of the field equations has a fully different meaning of that one it has within General Relativity theory. In the latter case, the trace of the field equations is a mere algebraic relation involving the Ricci scalar $R$ and the trace of the energy-momentum tensor $T$. On the other hand, in the context of a $f(R)$ gravity in the metric formalism, the trace of the corresponding field equations, given by the equation (\ref{trace}) is a dynamical equation, scalar field-like, with $f'=f'(R)$ playing the important role of the scalar field emerging as a consequence of the modified gravity. This scalar field is commonly called {\it scalaron} \cite{STAR}. In other words, this is a further scalar degree of freedom (propagating across the entire bulk), which the modification of the General Relativity theory gives rise. The equation (\ref{trace}) can be written as
\begin{equation}
\label{trace1}
\Box f'-\left(\frac{5f}{8}-\frac{f'R}{4}\right)=\frac{1}{4}k_{5}^{2}T\ .
\end{equation}
Let us notice that such an equation has the usual form of a Klein-Gordon-type equation for a scalar field $\phi$, which is
\begin{equation}
\Box \phi-\frac{dV}{d\phi}=S\ ,
\end{equation}
where $V$ corresponds to a potential associated with the scalar field and $S$ is a source term. Let us recall that the mass of the scalar field is obtained through the second derivative of the potential evaluated at a minimum of the field
\begin{equation}
\label{mass}
m_{\phi}^2=\left.\frac{d^2V(\phi)}{d\phi^2}\right|_{\phi=\phi_{min}}\ .
\end{equation}

For the equation (\ref{trace1}) we have an effective potential, $V_{eff}$, associated with the scalaron, defined as:
\begin{equation}
\label{eff}
\frac{dV_{eff}}{df'}=\left(\frac{5f}{8}-\frac{f'R}{4}\right)\ .
\end{equation}
As previously performed by \cite{pogosian}, by means of such expression, we can obtain the mass of the scalaron, by considering that its minimum lies at high curvature regimes, where it is assumed that $\left|f''R\right|\ll1$. Since General Relativity succeeds to describe correctly the phenomena observed in the universe at high redshifts, when the high curvature approximation is important, it is reasonable to choose $f'=1$ as the minimal value of the scalaron. So, by adopting such approximation and using the equations (\ref{mass}) and (\ref{eff}) the scalaron mass will be
\begin{equation}
\label{mass1}
m_{f'}^2\approx\frac{3}{8}\frac{1}{f''}\ ,
\end{equation}
which imposes positiveness for the second derivative of $f(R)$, i.e $f''>0$, in order to guarantee a tachyons-free theory. This is the first constraint to be satisfied by a viable $f(R)$ in our model.

\section{ The Appropriate Junction Conditions and the Projection Scheme}

The junction conditions `measure' (via its extrinsic curvature) how a co-dimension one hypersurface is embedded into a given bulk. When the embedding space gravity is given by the General Relativity theory, the junction conditions are the usual Israel-Darmois matching \cite{I,D}. Here we shall derive the appropriate conditions for a bulk in the framework of a $f(R)$ gravity.

The equation (\ref{1}) may be recast into a more familiar form, reading
\be R_{AB}-\frac{1}{2}g_{AB}R=\tilde{T}\el{bulk}_{AB},\label{2} \ee where
\be \tilde{T}\el{bulk}_{AB}=\frac{1}{f'(R)}\Bigg[k_{5}\el{2}T_{AB}-\Bigg(\frac{1}{2}Rf'(R)-\frac{1}{2}f(R)+\Box f'(R)\Bigg)g_{AB}+\nabla_{A}\nabla_{B}f'(R)\Bigg].\label{3} \ee Following the standard approach, we start from the Gauss equation relating the brane curvature tensor $\bar{R}\el{\alpha}_{\beta\gamma\delta}$ to its higher-dimensional counterpart
\be \bar{R}\el{\alpha}_{\beta\gamma\delta}=R\el{A}_{BCD}h\el{\alpha}_{A}h\el{B}_{\beta}h\el{C}_{\gamma}
h\el{D}_{\delta}+K\el{\alpha}_{\gamma}K_{\beta\delta}-K\el{\alpha}_{\delta}K_{\beta\gamma},\label{4} \ee where $h_{\mu\nu}$ is the brane metric, related to the bulk metric by $g_{AB}=h_{AB}+n_{A}n_{B}$ being $n_{A}$ an ortonormal vector in the extra dimension direction and $K_{\mu\nu}=h\el{A}_{\mu}h\el{B}_{\nu}h\nabla_{A}n_{B}$ is the extrinsic curvature of the 3-brane. With the aid of Eqs. (\ref{4}) and (\ref{3}) it is simple to obtain the Einstein tensor on the brane
\be \bar{G}_{\mu\nu}=\frac{2}{3}\Bigg[\tilde{T}\el{bulk}_{AB}h\el{A}_{\mu}h\el{B}_{\nu}+\Bigg(\tilde{T}\el{bulk}_{AB}n\el{A}n\el{B}-\frac{\tilde{T}\el{bulk}}{4}\Bigg)h_{\mu\nu}\Bigg]  +KK_{\mu\nu}-K_{\mu}\el{\sigma}K_{\sigma\nu}-\frac{1}{2}h_{\mu\nu}(K\el{2}-K\el{\alpha\beta}K_{\alpha\beta})-E_{\mu\nu},\label{5}\ee where $E_{\mu\nu}=C\el{A}_{BCD}n_{A}n\el{C}h\el{B}_{\mu}h\el{D}_{\nu}$, the usual projection of the $5D$ Weyl tensor $(C\el{A}_{BCD})$ encoding genuine extra-dimensional gravitational effects.

It is conceivable, for the projection scheme purposes, to decompose the bulk stress tensor into $T_{AB}=-\Lambda g_{AB}+S_{AB}\alpha\delta(y)$, where $\Lambda$ is the five dimensional cosmological constant, $S_{AB}$ the stress tensor on the brane, the Dirac distribution localizes the brane along the extra dimension\footnote{We note, by passing, that the functional form of $T_{AB}$ could also be complemented via the addition of a bulk matter. In particular, the junction conditions to be derived in what follows allow one to do so.} and $\alpha$ is a constant parameter with dimension of length, introduced here in order to allow $S_{AB}$ to have the same dimension of $T_{AB}$. Moreover, the brane energy-momentum tensor is also decomposed in a similar fashion, $S_{\mu\nu}=-\lambda h_{\mu\nu}+\tau_{\mu\nu}$, in which $\lambda$ is the brane tension and $\tau_{\mu\nu}$ stands for matter on the brane. From Eq. (\ref{3}) and the above decomposition of the stress-tensors it is easy to see that the first three terms appearing in the right hand side of Eq. (\ref{5}) are
\ba
\xi_{\mu\nu} &\equiv& \left. \tilde{T}\el{bulk}_{AB}h\el{A}_{\mu}h\el{B}_{\nu}+\tilde{T}\el{bulk}_{AB}n\el{A}n\el{B}h_{\mu\nu}-\frac{1}{4}\tilde{T}\el{bulk}h_{\mu\nu}=-\frac{1}{f'(R)}\Bigg[2k_{5}\el{2}\Lambda+(Rf'(R)-f(R))+2\Box f'(R)\right.\nonumber\\&-&\left.\frac{5k_{5}\el{2}\Lambda f'(R)}{4}-\nabla_{A}\nabla_{B}f'(R)n\el{A}n\el{B}\Bigg]h_{\mu\nu}
+\frac{1}{f'(R)}\nabla_{A}\nabla_{B}f'(R)h_{\mu}\el{A}h_{\nu}\el{B},\right. \label{6}\ea in such way that
\be \bar{G}_{\mu\nu}=\frac{2}{3}\xi_{\mu\nu}+KK_{\mu\nu}-K_{\mu}\el{\sigma}K_{\sigma\nu}-
\frac{1}{2}h_{\mu\nu}(K\el{2}-K\el{\alpha\beta}K_{\alpha\beta})-E_{\mu\nu}.\label{7} \ee

So far, the obtained equations agree with Sec. II of Ref. \cite{IR}. However, from now on, we shall generalize the usual junction conditions in order to accomplish the $f(R)$ bulk gravity and determine the extrinsic curvature terms in the projection procedure. In general grounds, we are assuming that the normal derivative of $K_{\mu\nu}$ may become large when compared with its variations along the brane dimensions \cite{RBBC}. Therefore, the relevant discontinuity is computed from $[K_{\mu\nu}]=K_{\mu\nu}\el{+}-K_{\mu\nu}\el{-}$, where $K_{\mu\nu}\el{\pm}$ means the limit of $K_{\mu\nu}$ approaching the brane from the $\pm$ side. These considerations can be made more precise by the following construction \cite{CEM}: consider the brane as a timelike surface intersected orthogonally by geodesics. In a given coordinate system it is possible to set the brane at $y=0$ in agreement with the previous decomposition of the bulk stress tensor. Hence it is possible to write $n_{A}=\partial_{A}y$. Moreover, it is quite convenient to define another coordinate system, say $z\el{\mu}$, installed upon the brane such that the hypersurface may be parametrized as $x\el{A}=x\el{A}(z\el{\mu})$. In this vein, it is possible to define the soldering jacobian $e\el{A}_{\mu}=\partial x\el{A}/\partial z\el{\mu}$, tangent to the curves belonging to the brane. Note that, by construction,  $e\el{A}_{\mu}n_{A}=0$ and thus, $e\el{A}_{\mu}$ shall act as a projector onto the brane directions. Now, it is well known that, with the help of the Heaviside distribution $\Theta(y)$, the bulk metric may be decomposed into different metrics on both sides of the infinitely thin brane by \cite{MC} $g_{AB}=\Theta(y)g_{AB}\el{+}+\Theta(-y)g_{AB}\el{-}$. The terms $g_{AB}\el{\pm}$ mean the metric in the $\pm$ side and the $\Theta(y)$ distribution obeys the rules  $\Theta\el{2}(y)=\Theta(y)$, $\Theta(y)\Theta(-y)=0$ and $d\Theta(y)/dy=\delta(y)$, which makes the calculations quite manageable.

By decomposing the metric via the Heaviside distribution all the relevant geometrical quantities may be obtained. For instance, the first derivative of the metric reads $g_{AB,C}=\Theta(y)g\el{+}_{AB,C}+\Theta(-y)g\el{-}_{AB,C}+\delta(y)[g_{AB}]n_{C}$, leading to the imposition of the Darmois junction condition $[g_{AB}]=0$, since the product $\Theta(y)\delta(y)$ (which would appear in the connection) is not defined in the distributional calculus. It is important to note that the Darmois condition guarantees the continuity of its tangential derivatives. Therefore the only possible discontinuity of the metric derivative shall be along the extra dimension: $[g_{AB,C}]=\kappa_{AB} n_{C}$, where the $\kappa_{AB}$ tensor will not appear in the final condition.

Following this line it is possible to connect the delta part of the Einstein tensor in the brane with the delta part of the decomposition $\tilde{T}_{AB}\el{bulk}=\Theta(y)\tilde{T}\el{+}_{AB}+\Theta(-y)\tilde{T}\el{-}_{AB}+\delta(y)\tilde{T}_{AB}$ \cite{MC}. Now, by construction, we have $\tilde{T}_{AB}\el{bulk}e\el{A}_{\mu}e\el{B}_{\nu}=\tilde{T}_{\mu\nu}$ in such a way that the Eq. (\ref{3}) reads
\be \tilde{T}_{\mu\nu}=\frac{1}{f'(R)}\Bigg[k_{5}\el{2}S_{\mu\nu}-\Bigg(\frac{1}{2}Rf'(R)-\frac{1}{2}f(R)+\Box f'(R)\Bigg)h_{\mu\nu}+e\el{A}_{\mu}e\el{B}_{\nu}\nabla_{A}\nabla_{B}f'(R)\Bigg].\label{pr1}\ee Note that in this last equation the first term is given by $S_{\mu\nu}$ (the stress tensor on the brane) and not by $T_{\mu\nu}$, since one must respect the constraint $\tilde{T}_{\mu\nu}n\el{\nu}=0$. Now, since the geometrical part of the Israel matching condition is not modified, the delta part of the Einstein tensor will be the same of the standard case \cite{MC}. Therefore, the appropriate junction condition reads straightforwardly
\be [K_{\mu\nu}]=-\alpha\Bigg(\tilde{T}_{\mu\nu}-\frac{1}{3}h_{\mu\nu}\tilde{T}\Bigg),\label{pr3}\ee with $\tilde{T}_{\mu\nu}$ given by Eq. (\ref{pr1}). Hence, it is possible to express the jump of the extrinsic curvature in terms of the brane stress tensor plus corrections coming from the $f(R)$ bulk. It is given by
\be [K_{\mu\nu}]=-\frac{\alpha}{f'(R)}\Bigg[k_{5}\el{2}\Big(S_{\mu\nu}-\frac{1}{3}h_{\mu\nu}S\Big)+\frac{1}{6}h_{\mu\nu}[Rf'(R)-f(R)]+e\el{A}_{\mu}e\el{B}_{\nu}\nabla_{A}\nabla_{B}f'(R)\Bigg],\label{pr4} \ee whose trace reads \be [K]=\frac{\alpha}{3f'(R)}\Big[k_{5}\el{2}S-2Rf'(R)+2f(R)-3\Box f'(R)\Big].\label{pr5}\ee
Note that by taking $f(R)=R$ ($f'(R)=1$) we recover the usual General Relativity case, as expected.

From Eqs. (\ref{pr3}) and (\ref{pr4}) it is possible to complete the projection procedure, writing down Eq. (\ref{7}) on the brane. In order to implement it we notice that the $Z_{2}$ symmetry imposes the condition  $n\el{+}_{A}\rightarrow -n\el{-}_{A}$, leading to $K\el{+}_{\mu\nu}\rightarrow -K\el{-}_{\mu\nu}$ \cite{ROY}. Therefore, it is possible to calculate the extrinsic curvature terms of Eq. (\ref{7}) on the brane (henceforth we suppress the label $\pm$, since the dependence of $K_{\mu\nu}$ in Eq. (\ref{7}) is  quadratic). The calculation of the extrinsic curvature terms is easy but tedious. We will suppress some details and furnish the main results in terms of the brane stress tensor. They are
\ba  KK_{\mu\nu}&=&\left. -\frac{\alpha^2}{12[f'(R)]\el{2}} \Bigg\{k_{5}\el{4}SS_{\mu\nu}-\frac{k_{5}\el{4}}{3}h_{\mu\nu}S\el{2}+\frac{5k_{5}\el{2}}{6}h_{\mu\nu}S[Rf'(R)-f(R)]+k_{5}\el{2}Se_{\mu}\el{A}e_{\nu}\el{B}\nabla_{A}\nabla_{B}f'(R)\right.\nonumber\\&-&\left. 2k_{5}\el{2}S_{\mu\nu}[Rf'(R)-f(R)]-\frac{1}{3}h_{\mu\nu}[Rf'(R)-f(R)]\el{2}-2[Rf'(R)-f(R)]e_{\mu}\el{A}e_{\nu}\el{B}\nabla_{A}\nabla_{B}f'(R)\right.\nonumber\\&-&\left. 3k_{5}\el{2}S_{\mu\nu}\Box f'(R)+k_{5}\el{2}h_{\mu\nu}S\Box f'(R)-\frac{1}{2}h_{\mu\nu}[Rf'(R)-f(R)]\Box f'(R)\right.\nonumber\\&-&\left.3\Box f'(R)e_{\mu}\el{A}e_{\nu}\el{B}\nabla_{A}\nabla_{B}f'(R) \Bigg\},\right.\label{pr6}\ea
\ba K_{\mu}\el{\sigma}K_{\nu\sigma}&=&\left.\frac{\alpha^2}{4f'(R)} \Bigg\{k_{5}\el{4}\Bigg(S_{\mu}\el{\sigma}S_{\nu\sigma}-\frac{2}{3}SS_{\mu\nu}+\frac{1}{9}h_{\mu\nu}S\el{2}\Bigg)+\frac{k_{5}\el{2}}{3}S_{\mu\nu}[Rf'(R)-f(R)]-\frac{k_{5}\el{2}}{9}Sh_{\mu\nu}[Rf'(R)-f(R)]\right.\nonumber\\&+&\left. k_{5}\el{2}(S_{\mu}\el{\sigma} e_{\nu}\el{A}+S_{\nu}\el{\sigma} e_{\mu}\el{A})e_{\sigma}\el{B}\nabla_{A}\nabla_{B}f'(R)-\frac{2}{3}k_{5}\el{2}Se_{\mu}\el{A}e_{\nu}\el{B}\nabla_{A}\nabla_{B}f'(R)+\frac{1}{36}h_{\mu\nu}[Rf'(R)-f(R)]\el{2}\right.\nonumber\\&+&\left.\frac{1}{3}[Rf'(R)-f(R)]e_{\mu}\el{A}e_{\nu}\el{B}\nabla_{A}\nabla_{B}f'(R)+e_{\mu}\el{A}e_{\nu}\el{B}\nabla\el{C}\nabla_{A}f'(R)\nabla_{C}\nabla_{B}f'(R)\Bigg\},\right.\label{pr7}\ea
\ba K\el{2}&=&\left.\frac{\alpha^2}{36f'(R)}\Bigg\{k_{5}\el{4}S\el{2}-4k_{5}\el{2}S[Rf'(R)-f(R)]-6k_{5}\el{2}S\Box f'(R)+12\Box f'(R)[Rf'(R)-f(R)]\right.\nonumber\\&+&\left.4[Rf'(R)-f(R)]\el{2}+9(\Box f'(R))\el{2}\Bigg\} \right.\label{pr8}\ea and
\ba K\el{\alpha\beta}K_{\alpha\beta}&=&\left. \frac{\alpha^2}{4f'(R)}\Bigg\{k_{5}\el{4}\Bigg(S\el{\alpha\beta}S_{\alpha\beta}-\frac{2}{9}S\el{2}\Bigg)-\frac{k_{5}\el{2}}{3}S[Rf'(R)-f(R)]+2k_{5}\el{2}S\el{\alpha\beta} e_{\alpha}\el{A}e_{\beta}\el{B}\nabla_{A}\nabla_{B}f'(R)\right.\nonumber\\&-&\left. \frac{2k_{5}\el{2}}{3}S\Box f'(R)+\frac{1}{9}[Rf'(R)-f(R)]\el{2}+\frac{1}{3}[Rf'(R)-f(R)]\Box f'(R)\right.\nonumber\\&+&\left.\nabla\el{A}\nabla\el{B}f'(R)\nabla_{A}\nabla_{B}f'(R) \Bigg\}. \right.\label{pr9}\ea

Now, from Eqs. (\ref{pr8}) and (\ref{pr9}), it is possible to calculate the scalar part of the extrinsic curvature terms in Eq. (\ref{7}). Taking into account the decomposition $S_{\mu\nu}=-\lambda h_{\mu\nu}+\tau_{\mu\nu}$ we have
\ba K\el{2}-K\el{\alpha\beta}K_{\alpha\beta}&=&\left. \frac{\alpha^2}{4f'(R)}\Bigg\{k_{5}\el{4}\Bigg[\frac{\tau\el{2}}{3}-\frac{2\lambda\tau}{3}+\frac{4\lambda\el{2}}{3}-\tau\el{\alpha\beta}\tau_{\alpha\beta}\Bigg]- \frac{k_{5}\el{2}}{9}(\tau-4\lambda)[Rf'(R)-f(R)]+(\Box f'(R))\el{2}\right.\nonumber\\&+&\left.\Box f'(R)[Rf'(R)-f(R)]+\frac{1}{3}[Rf'(R)-f(R)]\el{2}-2k_{5}\el{2}\tau\el{\alpha\beta} e_{\alpha}\el{A}e_{\beta}\el{B}\nabla_{A}\nabla_{B}f'(R)\right.\nonumber\\&-&\left. \nabla\el{A}\nabla\el{B}f'(R)\nabla_{A}\nabla_{B}f'(R) \Bigg\} .\right.\label{pr10}\ea In a similar fashion, the remaining extrinsic curvature terms of Eq. (\ref{7}) are given by
\ba KK_{\mu\nu}-K_{\mu}\el{\sigma}K_{\nu\sigma}&=&\left. -\frac{\alpha^2}{4f'(R)}\Bigg\{k_{5}\el{4}\Bigg(\tau_{\mu}\el{\sigma}\tau_{\nu\sigma}+\frac{1}{3}\tau\tau_{\mu\nu}\Bigg)-\frac{1}{3}k_{5}\el{2}\tau_{\mu\nu}\Big(2k_{5}\el{2}\lambda+[Rf'(R)-f(R)]+3\Box f'(R)\Big)\right.\nonumber\\&-&\left. \frac{1}{3}e_{\mu}\el{A}e_{\nu}\el{B}\nabla_{A}\nabla_{B}f'(R)\Big(k_{5}\el{2}\tau+2k_{5}\el{2}\lambda +[Rf'(R)-f(R)]+3\Box f'(R) \Big)+\frac{h_{\mu\nu}}{3}\Bigg[k_{5}\el{4}\tau\lambda-k_{5}\el{4}\lambda\el{2}
\right.\nonumber\\&+&\left. \frac{k_{5}\el{2}}{2}\tau [Rf'(R)-f(R)]-k_{5}\el{2}\lambda [Rf'(R)-f(R)]-\frac{1}{4}[Rf'(R)-f(R)]\el{2}\right.\nonumber\\&-&\left.k_{5}\el{2}\lambda \Box f'(R)+k_{5}\el{2}\tau \Box f'(R)- \frac{1}{2}[Rf'(R)-f(R)] \Box f'(R)\Bigg]\right.\nonumber\\&+&\left.k_{5}\el{2}\Big(\tau_{\mu}\el{\sigma}e_{\nu}\el{A}+\tau_{\nu}\el{\sigma}e_{\mu}\el{A}\Big)e_{\sigma}\el{B}\nabla_{A}\nabla_{B}f'(R)+e_{\mu}\el{A}e_{\nu}\el{B}\nabla\el{C}\nabla_{A}f'(R)\nabla_{C}\nabla_{B}f'(R)\Bigg\}. \right.\label{pr11}\ea

The last two equations allow one to bound all the extrinsic curvature terms in Eq. (\ref{7}) by means of the brane stress tensor components, as well as $f(R)$ and its derivative terms. We shall write it explicitly, since these terms encode the junction condition departure from the usual case. Hence, organizing similar terms for convenience we have
\ba KK_{\mu\nu}&-&\left.K_{\mu}\el{\sigma}K_{\nu\sigma}-\frac{1}{2}\Big(K\el{2}-K\el{\alpha\beta}K_{\alpha\beta}\Big)=-\frac{\alpha^2}{4f'(R)}\Bigg\{k_{5}\el{4}\Bigg(\tau_{\mu}\el{\sigma}\tau_{\nu\sigma}+\frac{1}{3}\tau\tau_{\mu\nu}+\frac{1}{6}\tau\el{2}-\frac{1}{2}h_{\mu\nu}\tau\el{\alpha\beta}\tau_{\alpha\beta}\Bigg)\right.\nonumber\\&-&\left. \frac{1}{3}k_{5}\el{2}\tau_{\mu\nu}\Big(2k_{5}\el{2}\tau_{\mu\nu}\lambda+[Rf'(R)-f(R)]+3\Box f'(R)\Big)-\frac{1}{3}e_{\mu}\el{A}e_{\nu}\el{B}\nabla_{A}\nabla_{B}f'(R)\Big(k_{5}\el{2}\tau+2k_{5}\el{2}\lambda\right.\nonumber\\&+&\left.[Rf'(R)-f(R)]+3\Box f'(R)\Big)+k_{5}\el{2}\Big(\tau_{\mu}\el{\sigma}e_{\nu}\el{A}+\tau_{\nu}\el{\sigma}e_{\mu}\el{A}\Big)e_{\sigma}\el{B}\nabla_{A}\nabla_{B}f'(R)+e_{\mu}\el{A}e_{\nu}\el{B}\nabla\el{C}\right.\nonumber\\&\times&\left.\nabla_{A}f'(R)\nabla_{C}\nabla_{B}f'(R)+\frac{1}{3}h_{\mu\nu}\Bigg[2k_{5}\el{4}\lambda\tau+ k_{5}\el{4}\lambda\el{2}+\frac{1}{3}k_{5}\el{2}\tau [Rf'(R)-f(R)]-\frac{1}{3}k_{5}\el{2}\lambda[Rf'(R)-f(R)]\right.\nonumber\\&+&\left.\frac{1}{4}[Rf'(R)-f(R)]\el{2}+[Rf'(R)-f(R)]\Box f'(R)- k_{5}\el{2}\lambda\Box f'(R)+k_{5}\el{2}\tau\Box f'(R)+\frac{3}{2}(\Box f'(R))\el{2}\right.\nonumber\\&-&\left.3k_{5}\el{2}\tau\el{\alpha\beta}e_{\alpha}\el{A}e_{\beta}\el{B}\nabla_{A}\nabla_{B}f'(R)-\frac{3}{2}\nabla\el{A}\nabla\el{B}f'(R)\nabla_{A}\nabla_{B}f'(R)\Bigg]\Bigg\}.\label{pr12} \right. \ea

Now it is possible to write down the effective gravitational field equation on the brane. Thus from eqs. (\ref{6}) and (\ref{pr12}), eq. (\ref{7}) reads
\ba \bar{G}_{\mu\nu}&=&\left. -\Lambda_{eff}h_{\mu\nu}+k_{5}\el{4}\pi_{\mu\nu}+k_{5}\el{2}G_{eff}\tau_{\mu\nu}+\frac{1}{f'(R)}\nabla_{A}\nabla_{B}f'(R)h_{\mu}\el{A}h_{\nu}\el{B}
+ G_{eff} e_{\mu}\el{A}e_{\nu}\el{B}\nabla_{A}\nabla_{B}f'(R)\right.\nonumber\\&-&\left.\frac{k_{5}\el{2}}{4[f'(R)]\el{2}}\Big(\tau_{\mu}\el{\sigma}e_{\nu}\el{A}+ \tau_{\nu}\el{\sigma}e_{\mu}\el{A}\Big)e_{\sigma}\el{B}\nabla_{A}\nabla_{B}f'(R)-\frac{1}{4[f'(R)]\el{2}}e_{\mu}\el{A}e_{\nu}\el{B}\nabla\el{C}\nabla_{A}f'(R)\nabla_{C}\nabla_{B}f'(R)\right.\nonumber\\&-&\left.E_{\mu\nu},\right. \label{pr13}\ea  where
\be G_{eff}=\frac{\alpha^2}{12[f'(R)]\el{2}}\Big[2k_{5}\el{2}\lambda+[Rf'(R)-f(R)]+3\Box f'(R)\Big],\label{pr14} \ee
\be \pi_{\mu\nu}=-\frac{\alpha^2}{4[f'(R)]\el{2}}\Bigg(\tau_{\mu}\el{\sigma}\tau_{\nu\sigma}+\frac{1}{3}\tau\tau_{\mu\nu}+\frac{1}{6}h_{\mu\nu}\tau\el{2}-\frac{1}{2}h_{\mu\nu}\tau\el{\alpha\beta}\tau_{\alpha\beta}   \Bigg)\label{pr15} \ee and
\ba \Lambda_{eff}&=&\left.\frac{\alpha^2}{4[f'(R)]\el{2}}\Bigg\{\frac{2}{3}k_{5}\el{4}\lambda\tau+\frac{1}{3}k_{5}\el{4}\lambda\el{2}+\frac{1}{9}k_{5}\el{2}\tau[Rf'(R)-f(R)]-\frac{1}{9}k_{5}\el{2}\lambda[Rf'(R)-f(R)]+\frac{1}{12}[Rf'(R)-f(R)]\el{2} \right.\nonumber\\&+&\left. \frac{1}{3} [Rf'(R)-f(R)]\Box f'(R)-\frac{k_{5}\el{2}}{3}\lambda\Box f'(R)+\frac{k_{5}\el{2}}{3}\tau\Box f'(R)+\frac{1}{2}(\Box f'(R))\el{2}-
k_{5}\el{2}\tau\el{\alpha\beta}e_{\alpha}\el{A}e_{\beta}\el{B}\nabla_{A}\nabla_{B}f'(R)\right.\nonumber\\&-&\left.\frac{1}{2}\nabla\el{A}\nabla\el{B}f'(R)\nabla_{A}\nabla_{B}f'(R) \Bigg\}+2k_{5}\el{2}\Lambda+[Rf'(R)-f(R)]+2\Box f'(R)-\frac{5k_{5}\el{2}\Lambda f'(R)}{4}\right.\nonumber\\&-&\left.\nabla_{A}\nabla_{B}f'(R)n\el{A}n\el{B}. \right.\label{pr16}\ea

In the next Section we shall implement a few important conditions coming from braneworld models. These constraints acts supplementing the covariant approach of this Section, leading to well defined five-dimensional braneworld models within $f(R)$ gravity. By now we just remark, by passing, an important output encoded in the  Eq. (\ref{pr14}). Note that the $f(R)$ bulk gravity is felt on the Newtonian effective constant. Thus, in order to guarantee a positive $G_{eff}$ it is necessary that\footnote{The brane tension $\lambda$ is the proper vacuum energy for isotropic branes. This reinforces the fact that $\lambda$ is a positive parameter \cite{AGH3}.}
\be 2k_{5}\el{2}\lambda+[Rf'(R)-f(R)]+3\Box f'(R)>0.\label{pr17} \ee Therefore Eq. (\ref{pr17}) is the first nontrivial constraint which must be respected by a gravitationally viable braneworld model in the $f(R)$ bulk context.

\section{Constraining and enabling $f(R)$ projected braneworlds}

In order to better appreciate the relevant constraints which have to be respected for a theoretically and physically  interesting model, we shall look at the simplest case of the vacuum on the brane, e. g.  $\tau_{\mu\nu}=0$. In this case, equations (\ref{pr13}) and (\ref{pr16}) give
\ba \bar{G}_{\mu\nu}&=&\left. -\Lambda_{eff}h_{\mu\nu}+\frac{1}{f'(R)}\nabla_{A}\nabla_{B}f'(R)h_{\mu}\el{A}h_{\nu}\el{B}
+ G_{eff} e_{\mu}\el{A}e_{\nu}\el{B}\nabla_{A}\nabla_{B}f'(R)\right.\nonumber\\&-&\left.\frac{1}{4[f'(R)]\el{2}}e_{\mu}\el{A}e_{\nu}\el{B}\nabla\el{C}\nabla_{A}f'(R)\nabla_{C}\nabla_{B}f'(R)-E_{\mu\nu},\right. \label{pr17}\ea  with
\ba \Lambda_{eff}&=&\left.\frac{\alpha^2}{4[f'(R)]\el{2}}\Bigg\{\frac{1}{3}k_{5}\el{4}\lambda\el{2}-\frac{1}{9}k_{5}\el{2}\lambda[Rf'(R)-f(R)]+
\frac{1}{12}[Rf'(R)-f(R)]\el{2}+\frac{1}{3} [Rf'(R)-f(R)]\Box f'(R)\right.\nonumber\\&-&\left.\frac{k_{5}\el{2}}{3}\lambda\Box f'(R)+\frac{1}{2}(\Box f'(R))\el{2}-
\frac{1}{2}\nabla\el{A}\nabla\el{B}f'(R)\nabla_{A}\nabla_{B}f'(R) \Bigg\}+2k_{5}\el{2}\Lambda+[Rf'(R)-f(R)]+2\Box f'(R)\right.\nonumber\\&-&\left.\frac{5k_{5}\el{2}\Lambda f'(R)}{4}\nabla_{A}\nabla_{B}f'(R)n\el{A}n\el{B}. \right.\label{pr18}\ea

In general, when dealing with $f(R)$ gravity based braneworld models, there are two types of physical inputs which may be accomplished by a well defined model: on the one hand, there are constraints coming from braneworld theory (such as the mentioned positivity of the brane tension) and, on the other hand, there are additional conditions arising from the requirement of a viable $f(R)$ gravity model, necessary to fit experimental and observational data.  Obviously, the implementation of the former type of conditions is indeed necessary. However, the direct use of the $f(R)$ constraints in our case is a naive approach, since these constraints are valid for usual four-dimensional $f(R)$ cosmological models and we are dealing with $f(R)$ gravity in the bulk. Besides, the obtained field equation ((\ref{pr17}), for instance) is quite different from the standard case.  In this vein, our approach here is to follow the clue of the braneworld constraints and complement it with some additional conditions appearing in the $f(R)$ theory, without making reference to any particular model.

In the final part of the Sec. II we mentioned that the brane tension should be positive; also, in order to get a attractive gravity on the brane we must impose that $G_{eff}>0$.
We can use equation (\ref{trace}) to remove $\Box f'(R)$ and express (\ref{pr17}) solely in terms of some constants, $f(R)$ and its derivatives with respect to $R$. In doing so we have
\begin{equation}
\label{pr19}
\frac{1}{4}Rf'+\frac{7}{8}f+\frac{15}{4}k_{5}^{2}|\Lambda|+2k_{5}^{2}\lambda>0\ .
\end{equation}

In a pursuit for a physically interesting braneworld model, it is conceivable to insist in having an AdS bulk. Therefore one shall write $\Lambda \rightarrow -|\Lambda|$. Besides, the minuteness of $\Lambda_{eff}$ (it shall be lower than $10\el{-120}$) is well known and may be associated to an important constraint.

Let us follow a similar procedure in (\ref{pr18}) by using (\ref{1}) and (\ref{2})  in order to remove the terms of derivative of $f(R)$ with respect of the coordinates, i.e  $\frac{1}{2}\nabla\el{A}\nabla\el{B}f'(R)\nabla_{A}\nabla_{B}f'(R)$, $\nabla_{A}\nabla_{B}f'(R)n\el{A}n\el{B}$ and $\Box f'(R)$ ending up with an equation depending only on $f(R)$ and its derivatives. With such an upper bound in the  Eq. (\ref{pr18}) the resulting constraint reads
\begin{eqnarray}
\label{pr20}
&&\frac{1}{3}k_{5}^{2}\lambda-\frac{1}{36}k_{5}^{2}\lambda R f'+\frac{5}{72}k_{5}^{2}\lambda f+\frac{1}{32}f^2+\frac{1}{6}k_{5}^{2}|\Lambda|f'R+\frac{5}{24}f k_{5}^{2} |\Lambda|+\frac{1}{8}f'^2R^2+\frac{5}{8}k_{5}^{2}|\Lambda|^2+\frac{1}{\alpha^2}f'^2k_{5}^{2}|\Lambda|\nonumber\\&+&\frac{3}{\alpha^2}f'^3R+\frac{1}{2}\frac{f'^2f}{\alpha^2}-\frac{5}{12}k_{5}^{2}\lambda|\Lambda|-\frac{1}{2}f'^2R_{AB}R^{AB}+\frac{5f'^^3k_{5}^{2}|\Lambda|}{\alpha^2}-\frac{4f'^3}{\alpha^2}R_{AB}n^{A}n^{B}=0\ .
\end{eqnarray}

The point to be stressed here is that any five-dimensional braneworld model based upon an AdS $f(R)$ bulk must respect Eqs. (\ref{pr19}) and (\ref{pr20}) in order to guarantee a well defined model from the gravitational point of view. Of course, Eqs. (\ref{pr19}) and (\ref{pr20}) are rather nontrivial and it is not expected that they could be satisfied by chance. We remark, by passing, that in the General Relativity limit none of the constraints imposes any bound on the curvature scalar, as expected.

While this paper addresses itself to obtain the effective gravitational equations on the 3-brane, coming from a five-dimensional $f(R)$ theory. In contrast to the previous work related to this issue \cite{IR}, we see that our results -- encoded in the equations (\ref{pr17}), (\ref{pr18}), and (\ref{pr14}) -- are quite different from those obtained in \cite{IR} (see equations (19)-(24)). The reason for such a discrepancy, as previously remarked, rests upon the fact that we generalize the junction conditions instead of using the same conditions of General Relativity. As shown in the end of \cite{IR}, that spherically symmetric solution may be used to explain the galaxy rotation curves, while some cosmological solution would be used to describe an accelerate universe. Equipped with the proper effective equations found here, we believe it is also possible, although very difficult, to find out suitable $f(R)$ models whose solutions in some regime describe the aforementioned behaviors. In fact, bearing in mind the essence of induced-gravity effects, it is expected in general grounds that at early times the usual cosmological behavior of the universe is restored, while at late times the standard results are no longer recovered and, for instance, acceleration can be driven by extra-dimensional gravity effects \cite{NASSO}.

We conclude this Section calling attention for a prominent difference encompassed by the equation (\ref{pr14}), coming from our approach, associated with a dynamical curvature. Roughly speaking, a time dependent scalar of curvature model would lead to a possible variation of the gravitational constant. Of course, the recent astrophysical data suggests the very constancy of the Newtonian gravitational constant. However, it is under current investigation its possible fractional variations \cite{WILL}. The best model independent bound on such a fractional variation is achieved by lunar ranging measurements establishing (4$\pm$ 9)$\times 10^{-13}$ $yr^{-1}$ \cite{LUNA}. Obviously it is a quite stringent constraint, but the point concerning our approach is that it could encompass such a variation.

\section{Final Remarks}

In this paper we have explored the viability of achieving a gravitational consistent braneworld scenario in the framework of a $f(R)$ theory. For this purpose, we have worked in the most general context without considering any particular model in order to derive the appropriate junction conditions to obtain a prescription on how to project the effective second order equations on a 3-brane embedded in a five-dimensional bulk. Besides providing a rigorous derivation of the junction conditions, our main result here is that any five-dimensional braneworld model based upon an AdS $f(R)$ bulk must respect Eqs. (\ref{pr19}) and (\ref{pr20}) in order to guarantee a well defined and gravitationally viable model.

Finally, let us give a brief exemplification of how to apply equations (\ref{pr19}) and (\ref{pr20}) to a particular (toy-)model. Let us consider a case in which we are dealing with a theory of gravity consisting of a small modification on General Relativity: $f(R)=R+\varphi(R)$, $\varphi(R)\ll R$. It is reasonable to single out models obeying this condition, since it is expected for an arbitrary $f(R)$ not to be so much different of General Relativity, which ensures the fulfilment of the viability conditions listed in \cite{pogosian}. By adopting such an assumption we will be led to the approximations below
\be f^2=R^2\left(1+\frac{2\varphi}{R}\right),\nonumber\ee
\be f'^2=1+2\varphi',\nonumber\ee
\be f'^3=1+3\varphi'\nonumber\ee and
\be f'^2f=R(1+2\varphi'+\frac{\varphi}{R}).\nonumber\ee
So, we will get a linear form for (\ref{pr20}) which reads
\begin{eqnarray}
\label{eqapp}
&&\left(a_{1}R^2+a_{2}R+a_{3}-R_{AB}R^{AB}-\frac{12}{\alpha^2}R_{AB}n^{A}n^{B}\right)\frac{d\varphi}{dR}+\left(b_{1}+b_{2}R\right)\varphi+c_{1}R^2+c_{2}R+c_{3}\nonumber\\&-&\frac{1}{2}R_{AB}R^{AB}-\frac{4}{\alpha^2}R_{AB}n^{A}n^{B}=0\ .
\end{eqnarray}
With the constants $a_1$, $a_2$, $a_3$, $b_1$, $b_2$, $c_1$, $c_2$ and $c_3$ given by
\begin{eqnarray}
\label{consts}
&&a_1\equiv\frac{1}{4}\ ,\;\;\;\;\;a_2\equiv\frac{1}{6}k_{5}^{2}|\Lambda|-\frac{1}{36}k_{5}^{2}\lambda+\frac{10}{\alpha^2}\ ,\;\;\;\;\;a_3\equiv\frac{17}{\alpha^2}|\Lambda|\ ,\;\;\;\;\;b_1\equiv\frac{1}{16}\ ,\;\;\;\;\;b_2\equiv\frac{5}{72}k_{5}^{2}\lambda+\frac{5}{24}k_{5}^{2}|\Lambda|+\frac{1}{2\alpha^2}\ ,\nonumber\\ &&c_1\equiv\frac{5}{32}\ ,\;\;\;\;\;\ c_2\equiv\frac{3}{72}k_{5}^{2}\lambda-\frac{3}{8}k_{5}^{2}|\Lambda|+\frac{7}{2\alpha^2}\ ,\;\;\;\;\;c_3\equiv\frac{1}{3}k_{5}^{2}\lambda+\frac{5}{8}k_{5}^{4}|\Lambda|^2+\frac{6k_{5}^{2}}{\alpha^2}|\Lambda|-\frac{5}{12}k_{5}^{2}\lambda|\Lambda|\ .
\end{eqnarray}
The equation (\ref{eqapp}) can just be solved in the case in which the scalars $R_{AB}R^{AB}$ and $R_{AB}n^An^B$ are  expressed as functions of the Ricci scalar $R$. Such equation represents an important condition of viability for models possessing the functional form $f(R)=R+\varphi(R)$ and obeying the constraint $\varphi(R)\ll R$. Apart of that, it is quite conceivable to implement a typical warped line element, expressing $R_{AB}R^{AB}$ and $R_{AB}n^An^B$ in terms of the warp factor. In this vein, the resulting differential equations for the warp factor in the coefficient of $d\varphi/dR$ and in the second term of Eq. (\ref{eqapp}) may be used to classify warped spaces accordingly to the correction $\varphi$.

\section*{Acknowledgments}
The authors thank to Centro Brasileiro de Pesquisas F\'{\i}sicas (CBPF/MCT) for hospitality during part of the development of this paper and to Profs. N. Deruelle,  J. A. Helay\"el-Neto and J. Lopes Neto for discussions during the preparation of this manuscript.  T. R. P. C., M. E. X. G. and J. M. H. S.  would like to thank CAPES  (Coordena\c{c}\~ao de Aperfei\c{c}oamento de Pessoal de N\'{\i}vel Superior) and CNPq (Conselho Nacional de Desenvolvimento Cient\'{\i}fico e Tecnol\'ogico), respectively, for financial support. Finally, the authors are grateful to the Referee for the important criticisms which helped to improve the previous version of this manuscript.

\end{document}